\documentclass[modern]{aastex61}

\usepackage{graphicx}
\usepackage{amsmath}
\usepackage{amssymb}
\usepackage{enumitem}

\newcommand\mybar{\kern1pt\rule[-\dp\strutbox]{.8pt}{\baselineskip}\kern1pt}

\setlist[itemize]{noitemsep, topsep=0pt, leftmargin=*}

\shorttitle{3I/ATLAS is Smaller or Rarer Than it Looks}
\shortauthors{Loeb}

\begin{document}

\title{Comment on "Discovery and Preliminary Characterization of a Third Interstellar Object: 3I/ATLAS" [arXiv:2507.02757]}

\author{Abraham Loeb}
\affiliation{Astronomy Department, Harvard University, 60 Garden
  St., Cambridge, MA 02138, USA}

\begin{abstract}
  The interstellar object 3I/ATLAS shows a weak cometary activity. Its
  brightness suggests a maximum radius of $\sim 10~{\rm
    km}~(A/0.05)^{-1/2}$ for an asteroid with an albedo $A$. I show
  that interstellar objects with that radius would amount to an
  interstellar mass density that is well above the expected mass
  budget of interstellar comets or asteroids. Given this budget, the
  detection rate of objects like 3I/ATLAS implies that it is a comet
  with a small core radius $<0.6~{\rm km}$, or a member of a rare
  population with a number density $\lesssim 5\times 10^{-8}~{\rm
    au^{-3}}$ for $R\gtrsim 10~{\rm km}$. The second possibility would
  suggest that the rare population of 3I/ATLAS objects favors plunging
  orbits towards the inner solar system to accommodate their inferred
  detection rate.
  
\end{abstract}

\section{Introduction}

The interstellar object
3I/ATLAS\footnote{https://www.minorplanetcenter.net/mpec/K25/K25N12.html}
was discovered on July 1, 2025 with an orbital eccentricity of $\sim
6.1$, perihelion of $\sim 1.36~{\rm au}$ (expected on October 29,
2025), inclination of $\sim 175^\circ$, and a hyperbolic velocity of
$\sim 58~{\rm km~s^{-1}}$~\citep{Seligman}.

3I/ATLAS shows a weak cometary activity. Its measured absolute
magnitude $H_V=12.4$ implies an upper limit on its nuclear radius of
$R\sim 10~{\rm km}(A/0.05)^{-1/2}$, for an asteroid-like albedo, $A$.
The detection rate of its population, $\sim 0.2~{\rm yr}^{-1}$,
requires a local number density of interstellar objects bigger than
3I/ATLAS, $n_0\sim 3\times 10^{-4}~{\rm au}^{-3}$, assuming random
orbits outside the Solar system~\citep{Seligman}.

Here, I show that the combination of these two estimates leads to an
untenable mass density in interstellar objects, unless 3I/ATLAS is a
comet with a much smaller core ($<0.6~{\rm km}$) or an object with
radius $\sim 10~{\rm km}$ that favored a plunging orbit into the inner
solar system as a member of a population with a much lower number
density ($\lesssim 5\times 10^{-8}~{\rm au^{-3}}$).

\section{Mass Density of Interstellar Objects}

The Galactic mass density of interstellar objects similar or bigger than
3I/ATLAS is given by,
\begin{equation}
\rho_{\rm ATLAS}= n_0 \times \left({4\pi\over 3}R^3\right)\times \rho_{\rm s}~,
  \end{equation}
where $\rho_{\rm s}$ is the intrinsic mean mass density of objects of
radius $>R$. Adopting $R$ as the radius of the solid core of 3I/ATLAS
and using the fiducial values inferred by \citet{Seligman} for $R$ and
$n_0$, I get:
\begin{equation}
\rho_{\rm ATLAS}= 0.01~M_\odot~{\rm pc}^{-3} \left({n_0\over 5\times
  10^{-4}~{\rm au^{-3}}}\right)\left({R\over 10~{\rm
    km}}\right)^3\left({\rho_{\rm s}\over 1~{\rm g~cm^{-3}}}\right)~.
  \label{massd}
  \end{equation}

For comparison, the Galactic mass density for stars in the
neighborhood of the Sun is $\rho_\star \sim 0.04 M_\odot {\rm
  pc}^{-3}$~\citep{McKee}. The total mass density in heavy elements
that sources interstellar comets or asteroids is $\rho_Z \sim
0.02\rho_\star\sim 8\times 10^{-4}~M_\odot~{\rm
  pc}^{-3}$. Furthermore, it is reasonable to expect the mass density
of interstellar asteroids to be lower than the mass density of rocky
materials around stars. Adopting an ejection to interstellar space of
$\sim 10 M_\oplus$ in comets or asteroids per solar mass in
stars~\citep{Loeb}, yields the limit
\begin{equation}
  \rho_{\rm ATLAS}\lesssim 10^{-6}~M_\odot~{\rm pc^{-3}}~.
\label{Upper}
\end{equation}

The limit from equations~(\ref{massd}) and ~(\ref{Upper}) implies that
either $R<0.6~{\rm km}$ for $n_0\sim 3\times 10^{-4}~{\rm au^{-3}}$ or
$n_0<5\times 10^{-8}~{\rm au^{-3}}$ for $R\sim 10~{\rm km}$. In the
first case, the radius limit from equations~(\ref{massd}) and
(\ref{Upper}) is about twice the limit on the core radius of the
interstellar comet 2I/Borisov~\citep{Jewitt}.

\section{Conclusions}

I conclude that the Galactic mass reservoir in ejected rocky materials
from planetary systems implies that either: {\it (i)} the reflection
of sunlight by 3I/ATLAS originates from its cometary plume whereas
most of its mass is contained in a compact solid core with a radius
$R\lesssim 0.6~{\rm km}$; or {\it (ii)} 3I/ATLAS contains a solid
object with a radius of $R\sim 10~{\rm km}$ but the local number
density of interstellar objects bigger than its size is limited to
$n_0<5\times 10^{-8}~{\rm au^{-3}}$. Both of the above possibilities
ease the tension of not detecting many more interstellar objects with
smaller radii than 3I/ATLAS, as sub-km objects are expected to be much
more numerous than $\sim 10~{\rm km}$ objects based on their
statistics in the solar system~\citep{Broz}.

If 3I/ATLAS is a comet, then it will get brighter as it comes closer
to the Sun and its surface gets warmer. Upcoming data from
state-of-the-art telescopes, including the {\it Vera C. Rubin
  Observatory} and the {\it James Webb Space Telescope}, will be able
to better constrain the nucleus radius of 3I/ATLAS in the coming
months. If 3I/ATLAS is a solid object with a physical radius $R\gtrsim
10~{\rm km}$, then the limited interstellar reservoir of rocky
materials would suggest that its trajectory favored a plunging orbit
towards the inner Solar system, perhaps by technological design.

\bigskip
\bigskip
\section*{Acknowledgements}

This work was supported in part by the Galileo Project and by
Harvard's {\it Black Hole Initiative}, which is funded by grants from
JFT and GBMF.

\bigskip
\bigskip
\bigskip

\bibliographystyle{aasjournal}
\bibliography{t}
\label{lastpage}
\end{document}